\documentclass[12pt,1p,times]{elsarticle}

\usepackage{graphicx} 
\usepackage{amssymb}
\usepackage{amsmath}
\usepackage{mathtools}
\usepackage{url}
\usepackage{xcolor}
\usepackage{pgf}
\usepackage{tikz}
\definecolor{InAGreen}{RGB}{173,195,60}

\journal{}
\begin{document}

\begin{frontmatter}



\title{Efficient low rank model order reduction of vibroacoustic problems under stochastic loads}


\author[InAD,SEEA]{Yannik Hüpel\corref{cor1}}
\cortext[cor1]{Corresponding author}
\ead{y.huepel@tu-braunschweig.de}
\author[InAD,SEEA]{Ulrich Römer}
\author[INUM]{Matthias Bollhöfer}
\author[InAD,SEEA]{Sabine C. Langer}

\affiliation[InAD]{organization={Institute for Acoustics and Dynamics, TU Braunschweig},
            addressline={Langer Kamp 19}, 
            city={Braunschweig},
            postcode={38106}, 
            country={Germany}}

\affiliation[INUM]{organization={Institute for Numerical Analysis, TU Braunschweig},
            addressline={Universitätsplatz 2}, 
            city={Braunschweig},
            postcode={38106}, 
            country={Germany}}

\affiliation[SEEA]{organization={Cluster of Excellence SE$^2$A–Sustainable and Energy-Efficient Aviation, TU Braunschweig},
            country={Germany}}

\begin{abstract}
This contribution combines a low-rank matrix approximation through Singular Value Decomposition (SVD) with second-order Krylov subspace-based Model Order Reduction (MOR), in order to efficiently propagate input uncertainties through a given vibroacoustic model. The vibroacoustic model consists of a plate coupled to a fluid into which the plate radiates sound due to a turbulent boundary layer excitation. This excitation is subject to uncertainties due to the stochastic nature of the turbulence and the computational cost of simulating the coupled problem with stochastic forcing is very high. The proposed method approximates the output uncertainties in an efficient way, by reducing the evaluation cost of the model in terms of DOFs and samples by using the factors of the SVD low-rank approximation directly as input for the MOR algorithm. Here, the covariance matrix of the vector of unknowns can efficiently be approximated with only a fraction of the original number of evaluations. Therefore, the approach is a promising step to further reducing the computational effort of large-scale vibroacoustic evaluations. 
\end{abstract}



\begin{keyword}
low rank approximation \sep model order reduction \sep turbulent boundary layer \sep vibroacoustics \sep stochastic modelling


\end{keyword}

\end{frontmatter}

\section{Introduction}
Aircraft cabin noise plays an important role in the design process of new aircraft. With rising passengers and larger climate impact the need for novel aircraft technologies is increasing rapidly \cite{EUROCONTROL.April2022}. However, when new technologies are developed and designed the novel aircraft design also has to be accepted by the passengers. Here, the noise comfort plays an important role. Generally, the higher the Sound Pressure Level (SPL) inside the aircraft cabin, the less likely passengers are to rate the flight as satisfactory \cite{Scheel.2016}. This induces the need for novel aircraft to have a reasonably low SPL inside the aircraft cabin. Therefore, vibroacoustic assessment of the cabin is a vital part in the aircraft design process. Naturally, this is also the case for other forms of mobility, while this contribution will focus on aircraft cabin noise as the demonstration example. 

Ever since the middle of the 20th century the vibroacoustic assessment of aircraft, due to changing excitation and materials, has been the subject of elaborate research \cite{WILBY1996545}. Especially when including this acoustic assessment in an early design stage, where usually no prototypes are present and the design includes many uncertainties such as material parameters, a simulative noise assessment becomes indispensable \cite{Blech.2022}. Here, so-called high-fidelity wave-resolving vibroacoustic models are needed to gain valuable insights into the acoustics of the aircraft cabin, seen by the steady increase of modelling complexity in relevant literature over the years \cite{buehrle,Herdic,MISSAOUI1999101}. Usually, vibroacoustic aircraft models are quite complex and of a large-scale, meaning they entail many Degrees of Freedom (DoFs) and solving the underlying system of equations can become computationally expensive \cite{buehrle,Roozen1992QuietBD}. If uncertainties in the excitation or the material properties need to be considered as well, the high computational effort makes an early design vibroacoustic assessment infeasible. 

The development of efficient numerical methods for multi-query tasks, such as the one that motivated the present study, has seen significant improvement in recent years. A common approach is to setup a surrogate model that, once constructed, can be efficiently evaluated many times. Here, polynomial methods, based for instance on polynomial chaos, have been proposed in \cite{sepahvand2012stochastic,sepahvand2017stochastic}. More recently, Gaussian process models and neural network models have also been investigated, see \cite{luegmair2024gaussian,van2023vibroacoustic} and \cite{cunha2023review} for a review of machine learning methods in vibroacoustics. 

In the field of vibroacoustics, Model Order Reduction (MOR) is another popular framework for surrogate modeling that can leverage the underlying structure of the equations \cite{ohayon2014advanced}. Both intrusive and non-intrusive MOR methods are in the scope of current research attempts. Efficient methods for frequency-dependent MOR have been proposed in \cite{aumann2023structured,xie2021efficient}. The case of combined parametric and frequency-dependent analysis has been considered from a multi-parameter perspective in \cite{rumpler2023mwcawe}, and in \cite{creixell2018adaptive} with a focus on optimization tasks and a hearing aid application. Furthermore, the inclusion of frequency dependent loads as well as multiple inputs and outputs have become easier to handle over the past years \cite{Sreekumar2021,Malh00,Hetm12}. Moreover, \cite{reyes2022multi,romeradaptive} have studied frequency-dependent MOR in combination with stochastic parameter variations. Despite the progress being achieved therein, taking all sources of uncertainty and large frequency ranges into account is still not possible for complex models and further advances in the computational methodologies are needed. 

In this work, we have access to the underlying finite element system of equations and therefore opt for an intrusive MOR approach. We demonstrate that one important source of uncertainty, represented by a frequency-dependent stochastic Turbulent Boundary Layer (TBL), can be handled efficiently by combining low-rank approximation and moment-matching model order reduction. We illustrate the efficacy of this new approach with a vibroacoustic benchmark, featuring a simple geometry. This example is well-suited to demonstrate the key characteristics of the method and the potentials in reducing computational effort and serves as an initial step towards the simulation of cabin noise considering complex sources of uncertainty. 

This paper is structured in the following way. We introduce the vibroacoustic and TBL models in Section \ref{sec:Model}. The MOR fundamentals are reported in Section \ref{sec:MOR}, while the new low-rank MOR method is introduced in Section \ref{sec:method}. Numerical results are given in Section \ref{sec:numerics} before some conclusions are drawn.

\section{Vibroacoustic model and set-up}
\label{sec:Model}
The following section introduces the model as well as the simulation set-up. 
The benchmark model consists of a rectangular plate that is excited by a TBL. Furthermore, the model includes a cavity into which the plate can radiate sound. Here, the simple benchmark model serves as an approximation to the previously stated aircraft cabin model. Through the excitation of the plate with the TBL, a cruise flight configuration is approximated. Therefore, the proposed model represents the essential mechanical properties with less computational effort compared to a complex vibroacoustic fuselage model. 

In order to spatially discretize the model, the Finite Element Method (FEM) is used. All of the computations are done in the frequency domain to avoid temporal discretization. 

\begin{figure}[h!]
    \centering
    \begin{tikzpicture}[scale=1]

\coordinate (A) at (0,0,0);
\coordinate (B) at (2,0,0);
\coordinate (C) at (2,2,0);
\coordinate (D) at (0,2,0);
\coordinate (E) at (0,0,2);
\coordinate (F) at (2,0,2);
\coordinate (G) at (2,2,2);
\coordinate (H) at (0,2,2);

\draw (A) -- (B);
\draw (B) -- (C);
\draw (C) -- (D);
\draw (D) -- (A);
\draw (E) -- (F);
\draw (F) -- (G);
\draw (G) -- (H);
\draw (H) -- (E);
\draw (A) -- (E);
\draw (B) -- (F);
\draw (C) -- (G);
\draw (D) -- (H);

\draw[fill={rgb,255:red,97;green,149;blue,171},opacity=0.5] (A) -- (B) -- (C) -- (D) -- cycle;
\draw[fill={rgb,255:red,97;green,149;blue,171},opacity=0.5] (E) -- (F) -- (G) -- (H) -- cycle;
\draw[fill={rgb,255:red,97;green,149;blue,171},opacity=0.5] (A) -- (B) -- (F) -- (E) -- cycle;
\draw[fill={rgb,255:red,97;green,149;blue,171},opacity=0.5] (C) -- (D) -- (H) -- (G) -- cycle;
\draw[fill={rgb,255:red,97;green,149;blue,171},opacity=0.5] (A) -- (D) -- (H) -- (E) -- cycle;
\draw[fill={rgb,255:red,97;green,149;blue,171},opacity=0.5] (B) -- (C) -- (G) -- (F) -- cycle;

\coordinate (R1) at (-2,0,0);
\coordinate (R2) at (-2,2,0);
\coordinate (R3) at (-2,2,2);
\coordinate (R4) at (-2,0,2);
\draw (R1) -- (R2);
\draw (R2) -- (R3);
\draw (R3) -- (R4);
\draw (R4) -- (R1);
\draw[fill=gray!40,opacity=0.75] (R1) -- (R2) -- (R3) -- (R4) -- cycle;

\coordinate (S1) at (-4,0,0);
\coordinate (S2) at (-4,2,0);
\coordinate (S3) at (-4,2,2);
\coordinate (S4) at (-4,0,2);
\draw (S1) -- (S2);
\draw (S2) -- (S3);
\draw (S3) -- (S4);
\draw (S4) -- (S1);
\draw[fill={rgb,255:red,67;green,67;blue,67},opacity=0.5] (S1) -- (S2) -- (S3) -- (S4) -- cycle;

\node[above] at (-2,2.6,0) {$\Gamma$ Interface};
\node[above] at (-4,2.5,0) {$\Omega_s$~Plate};
\node[above] at (1,2.5,0) {$\Omega_f$~Cavity};

\end{tikzpicture}
    \caption{Schematic of the examined model with an extruded view of the physical domain. $\Gamma$ denotes the fluid structure interface relevant for the strong coupling.}
    \label{fig:PlateCavSchematic}
\end{figure}
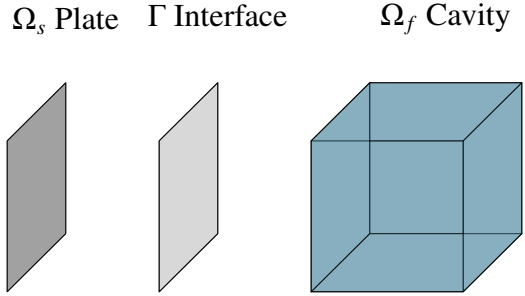

The examined model consists of two main domains that govern the mathematical equations used in this contribution. We have a structural domain $\Omega_s$ that is strongly coupled to the fluid domain $\Omega_f$ through the interface $\Gamma$. Therefore, the domains can be described by two Partial Differential Equations (PDEs) that are connected through a coupling condition. For the two domains the PDEs in frequency domain read
\begin{equation}
\begin{aligned}
    B\Delta\Delta u - \rho_s t \omega^2 u &= f, \quad &\text{in } \Omega_s, \\
    \Delta p + k^2p&=\rho_f\omega^2u, \quad &\text{in } \Omega_f,
\end{aligned}
\label{eq:PDEs}
\end{equation}
endowed with homogeneous Neumann boundary conditions. In Eq. \ref{eq:PDEs}, $k=\omega/c, B=(Et^3)/(12(1-\nu^2))$, where $B$ denotes the bending stiffness, with the material parameters of the structure being density $\rho_s$, thickness $t$, Young's modulus $E$ and Poisson's ratio $\nu$. The fluid's material properties are density $\rho_f$ and speed of sound $c$. The angular frequency is denoted by $\omega\in \mathbb{R}^+$. A strong coupling is ensured through boundary conditions at the coupling interface $\Gamma$, where the fluid can exert pressure on the plate and the plate's normal velocity excites the fluid. For a better visualization, Fig.~\ref{fig:PlateCavSchematic} highlights the two physical domains of the examined model and also showcases the coupling interface.

When spatially discretizing Eq.~\ref{eq:PDEs} we have to chose a element size such that the sound and bending waves are discretized with at least ten nodes per wavelength, preserving the wave-resolving nature of the FEM. Therefore, the spatial discretization is governed by the examined frequency domain and the material properties of the physical domains.
Finally, with Eq.~\ref{eq:PDEs}, applying a standard Galerkin FEM method we obtain the spatially discretized FEM equations
\begin{equation}
	\left[\mathbf{K}-\omega^2\mathbf{M}\right]\mathbf{x}(\omega,\theta)=\mathbf{f}(\omega,\theta),
	\label{eq:systemDis}
\end{equation}
where $\mathbf{K}, \mathbf{M} \in \mathbb{R}^{n \times n}$ denote the stiffness and mass matrix respectively. The frequency dependency is expressed through the angular frequency $\omega \in \mathbb{R}^+$. In Eq. \ref{eq:systemDis}, we account for uncertainty through an elementary random outcome $\theta$. Further, $\mathbf{x}(\omega,\theta) \in \mathbb{C}^n$ is the vector of unknowns that the system of equations has to be solved for and $\mathbf{f} \in \mathbb{C}^n$ the excitation containing the TBL contribution. Further details on the excitation will be given in the following section. As described above, the model contains a structural part and a fluid part. The global matrices in Eq. \ref{eq:systemDis} are given by the respective components
\begin{alignat}{4}
	\mathbf{K} &= 
	\begin{bmatrix}
		\mathbf{K}_s & -\mathbf{C}_\mathrm{sf} \\
		\mathbf{0}            & \mathbf{K}_f    \\
	\end{bmatrix} &&,~
	&\mathbf{M}&=
	\omega^2
	\begin{bmatrix}
		\mathbf{M}_s                                     & \mathbf{0}       \\
		\rho\mathbf{C}_\mathrm{sf}^\mathrm{T} & \mathbf{M}_f        \\
	\end{bmatrix} \label{eq:detailMat}, \\
	\mathbf{x}(\omega,\theta) &=
	\begin{bmatrix}
		\mathbf{u}(\omega,\theta) \\
		\mathbf{p} (\omega,\theta)\\
	\end{bmatrix} &&,~
 &\mathbf{f}(\omega,\theta) &=
	\begin{bmatrix}
		\mathbf{f}_\mathrm{TBL}(\omega,\theta) \\
		\mathbf{0} \\
	\end{bmatrix}, \label{eq:detailVec}
\end{alignat}
where the subscript denotes the respective fluid or structural part. $\mathbf{C}_\mathrm{sf}$ is the coupling matrix that couples the shared nodes and ensures a sound radiation from the plate into the cavity. Since the DoFs of plate and fluid have different physical interpretations $\mathbf{x}$ can be split into the plate's DoFs and the fluids sound pressure. The interpolated FEM solutions on the domain are denoted as $p_h,\mathbf{u}_h$, where $h$ denotes the maximum mesh size. Lastly, only the plate is excited due to the TBL and therefore only the plate's nodes are subjected to a non-zero RHS. 

\subsection{Turbulent Boundary Layer Excitation}
\label{sec:TBL}
In \cite{blech20jsv} a hybrid simulation chain was developed that allows for the inclusion of multiple sound sources in vibroacoustic cabin noise assessment. Here, the focus is laid on the TBL excitation, from which a RHS according to Eq. \ref{eq:systemDis} can be formed. The right hand side vector can be constructed with entries
\begin{equation}
    (\mathbf{f}_\text{TBL})_i =\int_{\Omega} p_h \psi_i \ d\Omega,
\end{equation}
where $\Omega = \bar{\Omega_s} \cup \bar{\Omega_f}$ and $\psi_i$ denotes an ansatz function for the pressure field. Therefore, in order to depict the excitation, the pressure underneath the TBL has to be computed, so that it can serve as input for the force vector assembly. This is done by assuming the excitation to be an elemental load. A workflow that allows for the inclusion of the TBL as noise source has been presented in \cite{Huepel.2023}. Additionally, the computation of the pressure underneath the TBL was adapted and presented in \cite{Huepel.2024} where the Uncorrelated Wall Plane Wave (UWPW) approach according to \cite{Maxit.2016} is used. It states that the pressure underneath a TBL can be computed as a superposition of of plane waves described by
\begin{equation}
    \mathbf{p}(\mathbf{x},\omega)=\sum_{h=1}^{N_x}\sum_{j=1}^{N_y}\sqrt{\frac{\Phi_{pp}(k_x^{(h)},k_y^{(j)},\omega)\delta k_x\delta k_y}{4\pi^2}}e^{i(k_x^{(h)} x+k_y^{(j)} y+\varphi_{hj})},
    \label{eq:TBL}
\end{equation}
where the exponential function describes a plane wave along the main dimensions of the plate, while the root term acts as an amplitude scaling to the plane wave. Here, $k_x^{(h)},k_y^{(j)}$ denote the considered wavenumbers in $x$ and $y$ direction of the plate and $\delta k_x\delta k_y$ denotes the wavenumber resolutions in both directions. 
The stochastic behavior of a TBL is modelled by adding a random phase $\varphi_{hj} \sim \mathcal{U}(0,2\pi)$ to the plane waves. 
Hence, $\mathbf{p}(\mathbf{x},\omega)=\mathbf{p}(\mathbf{x},\omega,\theta)$, which propagates to the RHS of the system of equations, which becomes a random vector. \\
In this contribution, a pressure field underneath the TBL from CFD data is created. We can evaluate a pressure excitation at any point of the CFD mesh. However, the nodes of the CFD mesh do not necessarily coincide with the nodes of the FEM mesh. Since both meshes are sufficiently small, a nearest neighbor interpolation is used to transfer the excitation to the FEM mesh.

\section{Model Order Reduction}
\label{sec:MOR}
The overall goal of MOR is to reduce the size of the system of equations by projecting the solution manifold onto a low-dimensional sub-manifold, while keeping the projection error small \cite{benner2015survey}.  Recent advances in the field of MOR have allowed to reduce more and more complex system in size.
There are many different MOR methods, but since we are mostly interested in the transfer functions of the dynamical vibroacoustic system a rational interpolation approach is the most promising for the second order system \cite{BAI2002,Bai2005}. These moment matching methods are particularly well-suited for wave-type problems and here we build on previous work \cite{sreekumar2021efficient,romeradaptive}. In moment-matching the full order space is reduced based on a generalized Krylov subspace of lower dimension generated by arbitrary $m\times n$ matrices $\mathbf{A}$, $\mathbf{B}$, and the starting vector $\mathbf{s}$ and the form \cite{Bai2005}
\begin{equation}
\begin{gathered}
    \mathcal{G(\mathbf{A},\mathbf{B},\mathbf{s})}_r=\text{span}\{\mathbf{s}_0,\mathbf{s}_1,...,\mathbf{s}_{r-1}\},\\   \mathbf{s}_0=\mathbf{s},\mathbf{s}_1=\mathbf{A}\mathbf{s}_0, \mathbf{s}_k=\mathbf{A}\mathbf{s}_{k-1}+\mathbf{B}\mathbf{s}_{k-2}~\text{for}~k\geq 2,
\end{gathered}
\end{equation}
where $r$ denotes the dimension of the Krylov subspace and the columnvectors spanning the subspace can efficiently be computed through moments \cite{benner2015survey}. The moments of the transfer function of the Full Order Model (FOM) can be matched as to create an adequate subspace of lower order, in which the accuracy of the reduced solution is still adequate. The actual reduction is done by projecting the FOM onto the subspace of lower order through a projection matrix \cite{benner2015survey,BennerFeng2021}.
With a fitting algorithm it becomes possible to compute a projection matrix $\mathbf{V} \in \mathbb{C}^{n\times r}$. With this matrix Eq.~\ref{eq:systemDis} can be rewritten according to
\begin{equation}
    \left[\mathbf{K}_r-\omega^2\mathbf{M}_r\right]\mathbf{x}_r(\omega,\theta)=\mathbf{f}_r(\omega,\theta),
	\label{eq:systemDisreduced}
\end{equation}
where the subscript $._r$ refers to reduced order, which can be obtained from
\begin{equation}
\begin{gathered}
    \mathbf{M}_r=\mathbf{V}^H\mathbf{M}\mathbf{V}, 
    \mathbf{K}_r=\mathbf{V}^H\mathbf{K}\mathbf{V}, \\
    \mathbf{x}_r=\mathbf{V}^H\mathbf{x},   
    \mathbf{f}_r=\mathbf{V}^H\mathbf{f}.  
\end{gathered}
\end{equation}
Here, we aim to construct a global projection matrix $\mathbf{V}$ that yields a valid reduced basis over the combined frequency and stochastic space and hence, is independent of $\omega,\theta$. It becomes obvious that after the projection matrix is obtained, only a system of size $r\times r$ has to be solved. Even if this smaller system contains dense matrices, if $r$ is significantly smaller than the original size of the system, the reduced system can typically be solved a lot faster and more efficiently, therefore saving computational effort. However, usually the computation of the projection matrix in the so-called offline phase takes some effort, meaning that the projection matrix should be computed only once and then used for each frequency step in order to gain an increase in efficiency. 

Here, we employ a rational Arnoldi algorithm to compute the basis of the Krylov subspace. In particular, an adaptive order rational Arnoldi algorithm is used to compute projection matrices that adhere to a certain error measure across the frequency domain of interest. Therefore, only the system matrices as well as the excitations have to provided, in order to obtain the projection matrix and to compute the Reduced Order Model (ROM). Details of the procedure can be found in \cite{Bodendiek.2014}.

\section{Combined Low-Rank MOR Method}
\label{sec:method}
Low-rank approaches have been combined with MOR procedures recently. For example, \cite{VANOPHEM201823,VANOPHEM2019597,Baur2014} successfully combined the two approaches for a vibroacoustic assessment. However, there the focus was laid on parametric changes for parametric MOR, while the proposed method in this contribution focuses to efficiently quantify the uncertainties due to the stochastic excitation. The inclusion of parametric uncertainties in this approach is subject of further research.\\
In \cite{Bui2008} the focus is laid on finding a reduced basis for a probabilistic aerodynamic problem, where again, the authors aim to express the system in a linear affine way, as to have a good input for the MOR procedure. This contribution differs by using the low-rank factors of the uncertain input directly as the input for the MOR procedure, therefore not needing a classical Monte-Carlo approach and also foregoing the affine description of the problem. \\
Other publications \cite{Carlberg2011,LIEU20065730,Amsall2008} successfully apply the related POD to their models as means of reducing the system size, however, these solely focus on reducing the CFD model and enhancing the computation thereof, while this contribution utilizes the CFD data as input without any influence on the model used for the aerodynamic computations. In the present paper, the novelty is the efficient quantification of uncertainty through combination of the low-rank and MOR method, together with the application to a vibroacoustic benchmark model. 

The overall goal of the simulation is to obtain displacement values for the excited plate and actual sound pressures for the cavity. With these results it will be possible to optimize the design to radiate less sound and consequently also to reduce the noise inside an aircraft cabin. Due to the excitation's stochastic nature, many evaluations of the model become necessary. If we consider a sample $\{\theta^{(i)}\}_{i=1}^{I}$ and denote with $K$ the number of frequency steps, $K \times I$ solutions of the FE-system are needed for an analysis of the stochastic TBL. For large-scale systems, the associated computational effort is prohibitive. 

The goal of a stochastic analysis is an approximation of the conditional density $\pi(\mathbf{x}|\omega)$. Here, we restrict ourselves to computing the mean vector and covariance of the solution from the mean and covariance of the right-hand-side. Please note that this would determine the sought density only in case of a normal distribution $\mathbf{x}(\omega,\cdot) \sim \mathcal{N}(\Bar{\mathbf{x}}(\omega),\boldsymbol{\Sigma}_x(\omega))$, where $\Bar{\mathbf{x}},\boldsymbol{\Sigma}_x$ denote the mean vector and covariance function, respectively. However, we emphasize that the method introduced below could also be used to approximate higher order moments and hence more complex distribution functions. 

\begin{figure}[h!]
    \centering
    \scalebox{0.8}{\input{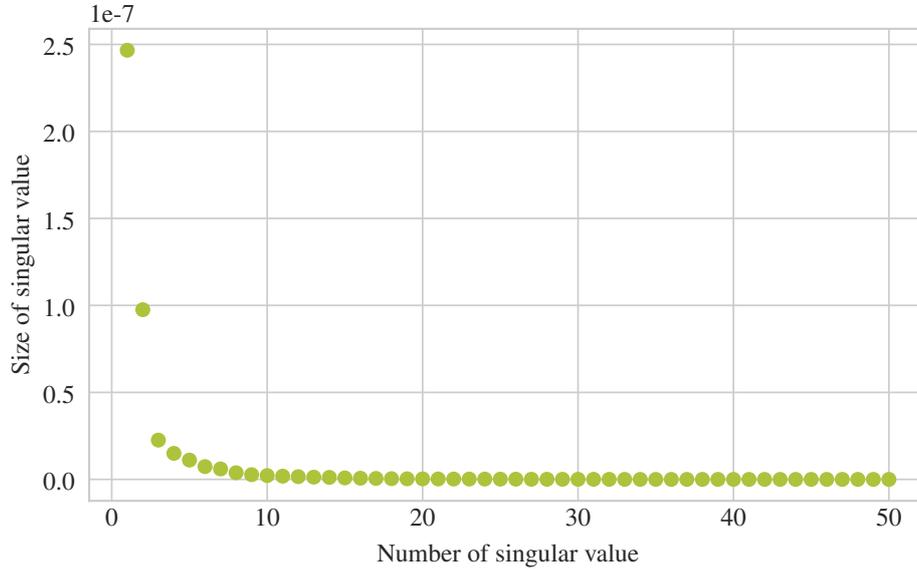}}
    \caption{Size of singular values of the covariance of the excitation. For simplicity, only the first $50$ singular values are shown here.}
    \label{fig:sv}
\end{figure}

We proceed by rewriting Eq.~\ref{eq:systemDis} more compactly as
\begin{equation}
	\mathbf{A}(\omega)\mathbf{x}(\omega,\theta)=\mathbf{f}(\omega,\theta).
	\label{eq:systemDis2}
\end{equation}
In view of the linearity of $\mathbf{A}$, the mean vector can be characterized by the system
\begin{equation}
	\mathbf{A}(\omega)\Bar{\mathbf{x}}(\omega)=\Bar{\mathbf{f}}(\omega),
	\label{eq:systemDis3}
\end{equation}
where $\Bar{\mathbf{f}}(\omega)$ denotes the mean excitation. Hence, for the solution mean one discrete vibroacoustics problem has to be solved over the frequency band of interest. 
There are several ways to compute the covariance matrix of the vector of unknowns $\mathbf{x}$, even though, they are numerically costly. The straight forward approach is to solve Eq.~\ref{eq:systemDis2} for each RHS sample and frequency step respectively and then estimate the covariance matrix for each frequency step. However, due to the computational effort, this is not feasible, but also not required. Again, because of the linearity of $\mathbf{A}$ the covariance can be obtained as 
\begin{equation}
     \boldsymbol{\Sigma}_x(\omega)= \mathbf{A}(\omega)^{-1} \boldsymbol{\Sigma}_f(\omega) \mathbf{A}(\omega)^{-H},
     \label{eq:covx}
\end{equation}
where $\boldsymbol{\Sigma}_f$ denotes the covariance matrix of the excitation. Eq.~\ref{eq:covx} showcases that for each frequency, a matrix equation has to be solved. Still, the equation requires the numerical inversion of $\mathbf{A}$, which is prohibitive for large-scale systems, such as the one examined in this contribution. In the literature, sparse approximation methods have been investigated, see for instance \cite{harbrecht2010finite}. Here, we simplify  Eq.~\ref{eq:covx} by utilizing the low-rank approximation of $\Sigma_f$ \cite{Eckart.1936}
\begin{equation}
    \boldsymbol{\Sigma}_f=\mathbf{U}\mathbf{S}\mathbf{V}^H\approx \mathbf{U}_l\mathbf{S}_l\mathbf{V}_l^H,
\end{equation}
where columns of $\mathbf{U}~\text{and}~\mathbf{V}$ indicate the left and right singular vectors, respectively, while $\mathbf{S}$ is the matrix of singular values. In general these matrices are complex and frequency-dependent, for brevity and better visualization we omit this in the following. The subscript $l$ denotes the low-rank approximation, meaning that only singular values up to rank $l$ are considered, automatically leading to a smaller size matrix. In the following the columns of $\mathbf{U}~\text{and}~\mathbf{V}$ are called low-rank factors. With these low-rank factors it becomes possible to efficiently compute $\Sigma_x$ by propagating the low-rank factors of $\mathbf{f}$ through the governing system of equations
\begin{equation}
    \boldsymbol{\Sigma}_x\approx \mathbf{U}_x\mathbf{S}_x\mathbf{V}_x^T,
\end{equation}
where $\mathbf{S}_x$ is obtained directly from the SVD of $\boldsymbol{\Sigma}_f$, while 
\begin{equation}
    \mathbf{A}\mathbf{U}_{x,i}=\mathbf{U}_i,
    \label{eq:covsvd}
\end{equation}
which works analogously for $\mathbf{V}_x$. This means, that in order to compute the covariance matrix of $\mathbf{x}$, Eq.~\ref{eq:covsvd} only has to be solved $l$ times for the left and right low-rank factors, respectively, ultimately requiring the system to be solved $2l$ times (in fact only $l$-times because left and right low-rank factors are the same for a symmetric covariance matrix). Hence, we create the low-rank approximation of $\boldsymbol{\Sigma}_x$ through the governing system of equations. 
One can find $l<n$, where
\begin{equation}
    \sum^{n}_{i=1}S_i\approx \sum^{l}_{i=1}S_i,
    \label{eq:criterion}
\end{equation}
if the singular values decrease rapidly, see also Fig.~\ref{fig:sv}.
It can be seen that there is a steep drop-off of the size of the singular values of the covariance matrix $\boldsymbol{\Sigma}_f$, meaning that a low-rank approximation is feasible. Ultimately, this already should decrease the required amount of system solutions per frequency, corresponding to a large reduction regarding the computational effort.\\
However, when looking at large-scale models, the computational cost can still be high. 
Therefore, we also apply a MOR approach to additionally reduce the size of the system to be solved, further increasing the efficiency of the examined approach. The previous section introduced the MOR approach used in this contribution, which allows for the reduction of DOFs. As mentioned in the previous section, the rational Arnoldi algorithm yields a projection matrix that allows us to reduce our full sized model to a smaller dimension. This algorithm achieves moment matching at a number of expansion points. Hence, at each expansion point we have to compute the low-rank factors. The collection of mean vectors and low-rank factors at every expansion point provides the set of input vectors for the Arnoldi algorithm. The resulting ROM model can then be solved at each frequency step in the online phase, at a reduced computational cost with regard to the FOM, as our numerical results will show. \\

\section{Numerical Results}
\label{sec:numerics}
\begin{figure}[h!]
    \centering
    \includegraphics{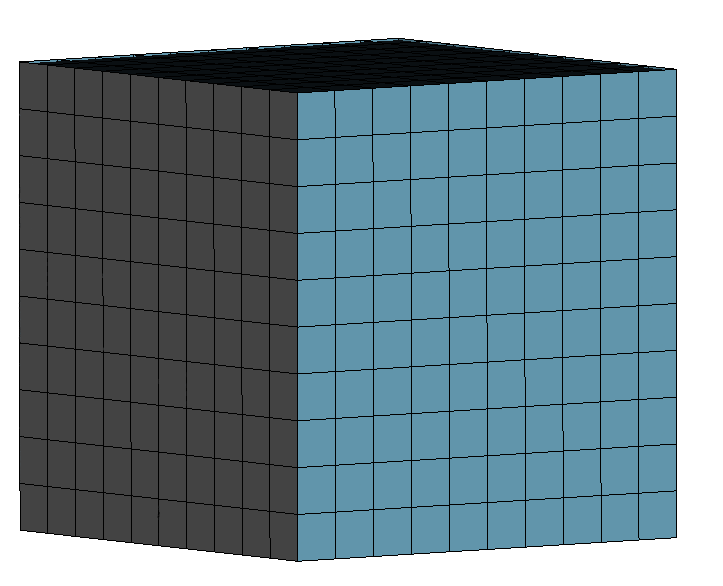}
    \caption{Plate (gray) and cavity (blue) model. The plate is excited by a TBL and radiates sound into the cavity.}
    \label{fig:PlateCav}
\end{figure}

Finally, with the methodology introduced in the previous section, the following section will focus on the application of the approach to examine the presented plate cavity model. First, results will be presented that showcase the accuracy of the approach, afterwards the computational effort will be discussed in terms of solving times.\\
The plate is modelled with nine-noded shell elements with six DOFs per node. Namely, those are the translational and rotational DOFs. 
Since only a general schematic of the examined model has been introduced, Figure \ref{fig:PlateCav} shows the spatially discretized model used in this contribution.
The model as shown above has a total of $40800$ DOFs, which showcases that even a simple looking model entails substantial computational effort, especially when keeping in mind, that many evaluations become necessary for uncertainty quantification. The plate is modelled with nine-node shell elements with six DOFs per node, while the fluid is modeled as $27$-node hexahedron with only the pressure as DOF. Furthermore, the considered frequency domain is $16-500$~Hz, with a resolution of $2$~Hz, where $16$~Hz marks the beginning of the audible range. The upper frequency limit is chosen to keep the spatial discretization fairly coarse, entailing less computational effort, while also obtaining interpretable results. The plate is made up of aluminum, while the cavity is modelled as air at room temperature. Table \ref{tab:matprops} shows the material properties of the model.
\begin{table}[h!]
    \centering
    \caption{Material properties of the plate cavity model.}
    \begin{tabular}{c|c}
         Parameter& Value \\
         \hline
         Young's Modulus E&$70$~GPa \\
         Thickness t&$0.003$~m \\
         Poisson's ratio $\nu$& $0.3$ \\
         Density plate $\rho_p$& $2700$~$\frac{\text{kg}}{\text{m}^3}$ \\
         Density air $\rho_f$& $1.21$~$\frac{\text{kg}}{\text{m}^3}$  \\
         Speed of sound c&$340$~$\frac{\text{m}}{\text{s}}$\\
    \end{tabular}
    \label{tab:matprops}
\end{table}

Additionally, some details about the TBL excitation are given here. The TBL implementation was carried out in python and for the random phase the numpy.random.rand() \cite{harris2020array} function was used. For this contribution, $1000$ samples are drawn, meaning we considered $1000$ different excitations per frequency step. Still, $40800$ DOFs entail a significant computational effort for each sample per frequency step and it becomes necessary to create efficient ROMs.
For the frequency domain examined here, two expansion points where chosen at $114 ~\text{and}~ 414$~Hz a priori. We found that this choice lead to sufficiently accurate results and therefore, no adaptive greedy selection of expansion points was used, which could further reduce the error. In order to now compute the projection matrix, the system matrices as well as the right hand side at the expansion points have to be supplied, respectively. 

As the overall goal is to approximate the mean and covariance of the vector of unknowns $\mathbf{x}$, the RHS chosen here are the mean of all excitations and the low-rank factors of the covariance $\boldsymbol{\Sigma}_f$. In theory, this should result in a ROM with which we can approximate the mean of $\mathbf{x}$ and the low-rank factors of $\boldsymbol{\Sigma}_x$, by solving the reduced version of Eq.~\ref{eq:systemDis2} and Eq.~\ref{eq:covsvd}. This approach therefore not only aims to reduce the dimensionality of the problem in the stochastic domain, but also to reduce the overall amount of DOFs. According to Figure \ref{fig:sv} and with the criterion from Eq.~\ref{eq:criterion}, $15$ low-rank factors for the right and left side are chosen, plus an additional RHS for the mean, ultimately resulting in $31$ inputs per expansion point. The final size of the ROM is determined by the number of inputs, the expansion points and the order. Here, with two expansion points and an order of $20$ it results in $1240$ DOFs, significantly reducing the amount of DOFs. 

\begin{figure}[h!]
\begin{tikzpicture}[>=latex]
  \draw[->, thick] (0,0,0) -- (3,0,0) node[anchor=north east] {$243$ frequency steps};
  \draw[->, thick] (0,0,0) -- (0,3,0);
  \draw[->, thick] (0,0,0) -- (0,0,-3);
  \node at (2.5, 2,-3)   (a) {$1001$ equations to solve};
  \node at (0, 3.1,0)   (c) {$40800$ DOFs};
  \draw[fill=InAGreen, fill opacity=0.5] (1.1,1,2.5) -- (1.1,2,2.5) -- (1.1,2,-2) -- (1.1,1,-2) -- cycle;
  \draw[fill=InAGreen, fill opacity=0.5] (1.6,1,2.5) -- (1.6,2,2.5) -- (1.6,2,-2) -- (1.6,1,-2) -- cycle;
  \draw[fill=InAGreen, fill opacity=0.5] (2.1,1,2.5) -- (2.1,2,2.5) -- (2.1,2,-2) -- (2.1,1,-2) -- cycle;
  \draw[fill=InAGreen, fill opacity=0.5] (2.6,1,2.5) -- (2.6,2,2.5) -- (2.6,2,-2) -- (2.6,1,-2) -- cycle;
  \draw[fill=InAGreen, fill opacity=0.5] (3.1,1,2.5) -- (3.1,2,2.5) -- (3.1,2,-2) -- (3.1,1,-2) -- cycle;
\end{tikzpicture}
\begin{tikzpicture}[>=latex]
  \begin{scope}[xshift=10cm]
    \draw[->, thick] (0,0,0) -- (3,0,0) node[anchor=north east] {$243$ frequency steps};
    \draw[->, thick] (0,0,0) -- (0,3,0) node[anchor=west] {$1240$ DOFs (MOR)};
    \draw[->, thick] (0,0,0) -- (0,0,-3) ;;
    \node at (2, 1,-3)   (b) {$31$ equations to solve (low-rank)};
  \draw[fill=InAGreen, fill opacity=0.5] (1.1,1,2.5) -- (1.1,1.75,2.5) -- (1.1,1.75,0) -- (1.1,1,0) -- cycle;
  \draw[fill=InAGreen, fill opacity=0.5] (1.6,1,2.5) -- (1.6,1.75,2.5) -- (1.6,1.75,0) -- (1.6,1,0) -- cycle;
  \draw[fill=InAGreen, fill opacity=0.5] (2.1,1,2.5) -- (2.1,1.75,2.5) -- (2.1,1.75,0) -- (2.1,1,0) -- cycle;
  \draw[fill=InAGreen, fill opacity=0.5] (2.6,1,2.5) -- (2.6,1.75,2.5) -- (2.6,1.75,0) -- (2.6,1,0) -- cycle;
  \draw[fill=InAGreen, fill opacity=0.5] (3.1,1,2.5) -- (3.1,1.75,2.5) -- (3.1,1.75,0) -- (3.1,1,0) -- cycle;
  \end{scope}
\end{tikzpicture}
\caption{Left: Starting point of this contribution. Right: After low-rank and MOR, significant reductions of dimensions and equations can be obtained.}
\label{fig:ReducedTikz}
\end{figure}
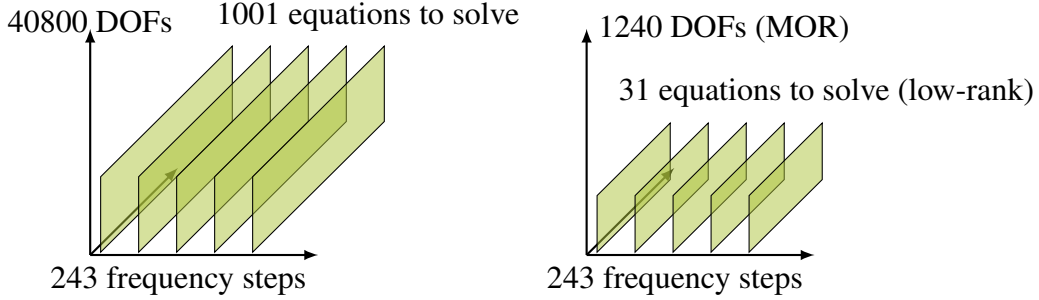

\begin{figure}[h!]
    \centering
    \scalebox{0.8}{\input{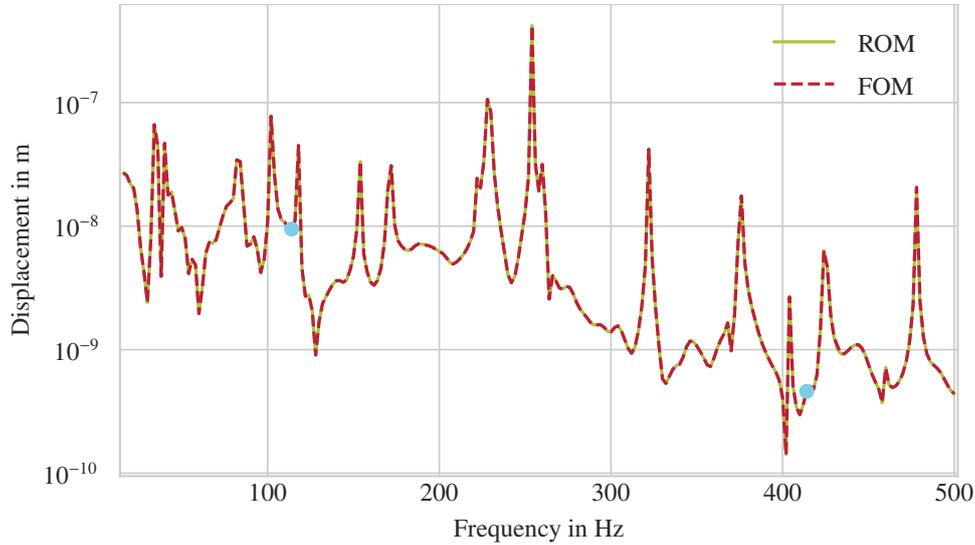}}
    \caption{Comparison between a displacement DOF of the FOM (red) and ROM (green). The expansion points are marked in blue.}
    \label{fig:PlateDofDis}
\end{figure}

\begin{figure}[h!]
    \centering
    \scalebox{0.8}{\input{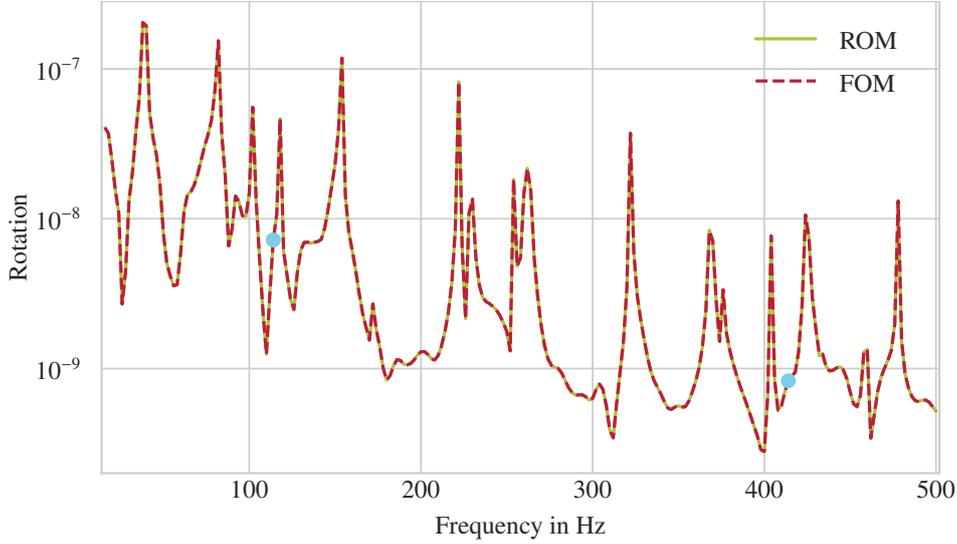}}
    \caption{Comparison between a rotation DOF of the FOM (red) and ROM (green). The expansion points are marked in blue.}
    \label{fig:PlateDofRot}
\end{figure}

\begin{figure}[h!]
    \centering
    \scalebox{0.8}{\input{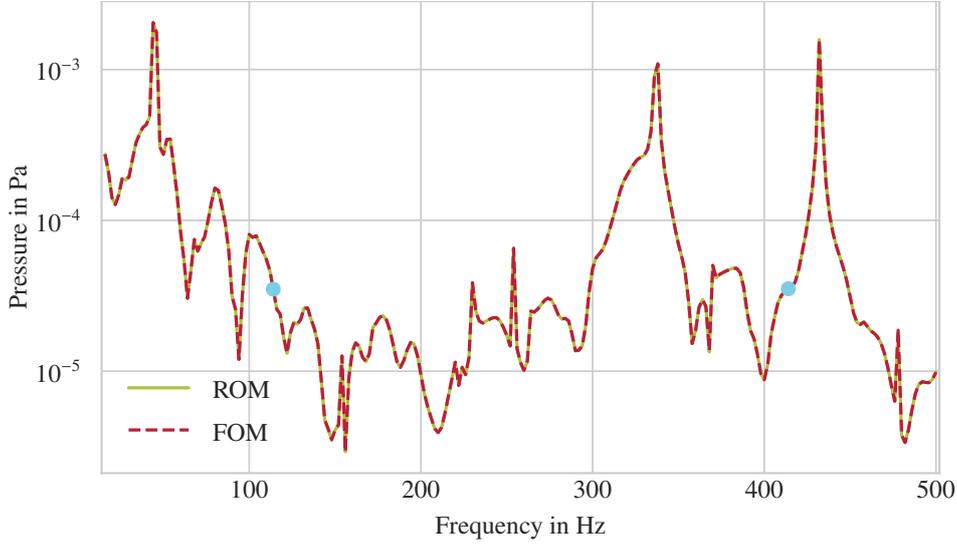}}
    \caption{Comparison between a pressure DOF of the FOM (red) and ROM (green). The expansion points are marked in blue.}
    \label{fig:FluidDof}
\end{figure}

To further showcase the reduction measures of this contribution, Fig.~\ref{fig:ReducedTikz} depicts the reductions due to the low-rank and MOR approach.
It can be seen that even though an extensive offline phase might be needed, the online phase can leverage a significantly reduced system size, which makes the approach promising. The green rectangles serve as indicators of the size of the system of equations with regard to DOFs and number of equations. When applying the proposed method we can reduce the DOFs with MOR, making one side of the rectangles smaller.
A ROM has to approximate all the DOFs well in order to be usable and adequate. Therefore, in order to evaluate the ROM, comparisons with different DOFs have to be made and the error across the frequency domain has to be computed, to also give a quantitative measure about the quality of the ROM. As a reminder, the overall goal of the ROM is to compute the mean of the vector of unknowns, as well as the covariance matrix per frequency step. Therefore, we first have to see if the mean is approximated correctly, after which we can evaluate the covariance matrices. For the computation of the mean Eq.~\ref{eq:systemDis3} is solved with a ROM and compared to the FOM solution. Fig.~\ref{fig:PlateDofDis} shows the comparison between an arbitrarily chosen node of the FOM and ROM. The DOF is the displacement in z-direction, which is an important DOF, since it induces the radiation into the cavity.
Just by examining Fig.~\ref{fig:PlateDofDis} it becomes obvious that for the shown DOF the ROM seems to approximate the mean response of the displacement in z-direction well. No major deviations from the FOM can be seen, meaning that the $1240$ DOF ROM is a good surrogate for the $40800$ DOF FOM. However, this is only a qualitative comparison, for a more meaningful interpretation we introduce a straight forward error measure that serves as an indicator of the approximation accuracy,
\begin{equation}
    \text{err}=\text{max}(||p_{\text{FOM}}|-|p_{\text{ROM}}||).
\end{equation}
This is an absolute error measure of the pressure across the whole frequency domain and serves as a quantitative indicator of the accuracy of the obtained ROM. In Figure \ref{fig:PlateDofDis} $\text{err}=8.45\times 10^{-10}$, which gives an relative error of $6\%$. This showcases the fact that the ROM is also accurate with regards to quantitative measures, meaning that at least for a displacement DOF the ROM can be used to estimate the FOM solution.\\
Still, there are other physical DOFs in the model that need to be approximated by the ROM. The shell elements also entail rotational DOFs, which are not that important for the radiation problem and ultimately for the sound pressure inside the cavity, but they should still be approximated well enough by the ROM. Therefore, Fig.~\ref{fig:PlateDofRot} highlights the comparison between the rotational DOF of the same node as above.
Just as in the previous figure, it can be seen that both plots seem to match perfectly across the examined frequency domain. The resonance are approximated well and even in-between the ROM estimates the FOM quite well, meaning that the ROM can be used in the computation of the mean. This is supported by the introduced error measure $\text{err}=5.71\times 10^{-10}$, which corresponds to a relative error of $4.5\%$. Again, it can be seen that error in the frequency domain is quite low. However, it should be mentioned here that with increasing frequency the error is expected to increase, because as the modal density increases, so does the error, since it gets more difficult for the ROM to approximate many modes in a high frequency regime. Still, the ROM delivers a fitting estimation in the frequency domain up to $500$~Hz. This means, that at least for a plate excited through a stochastic elemental load due to a TBL a ROM can be built and used for the approximation of the FOM.\\
However, in a vibroacoustic setting we are often more interested in the sound pressure field inside the cavity, e.g. inside an aircraft cabin. Therefore, we also have to evaluate the ROM's accuracy inside the fluid domain. Again, just as above, an arbitrary node is chosen to evaluate the comparison between FOM and ROM. Now, the sound pressure at the cavity node is compared and the result is shown in Fig.~\ref{fig:FluidDof}.
First of all, it can be seen that the overall magnitude of the values is multiple orders higher than the results of the plate, which is to be expected. These pressure values would be audible and therefore also have a physical meaning. More important, however, is the fact that the transfer functions of FOM and ROM again seem to match perfectly over the examined frequency domain. Again, this is supported by the introduced error measure $\text{err}=1.10\times 10^{-6}$, which is higher compared to the other DOFs, but when keeping in  mind that the magnitude of the results is also higher it is an acceptable error of $7.5\%$. 
Therefore, it can be said that the ROM estimates both the plate and the cavity well and can be used for the computation of the mean. 
Since the excitation is stochastic, the covariance matrix has to be computed for every frequency step. As mentioned in Sec.~\ref{sec:method}, the computation is computationally challenging, which is why we seek to combine the low-rank approximation approach with the MOR method already successfully applied to the computation of the mean. It is important to mention here that, through utilizing the low-rank factors as input for the MOR algorithm, we created one single projection matrix, that in an ideal case, approximates both the mean and the covariance factors adequately. However, a graphical evaluation of the covariances is non-trivial, since the matrices are of size $40800\times 40800$, which is already not tractable for for many computers. To still gain valuable insights into the prediction accuracy of the ROM with regards to the covariance matrix, we introduce another error measure, namely
\begin{equation}
    \text{err}_{\text{cov}}=\frac{||\Sigma_{x,\text{FOM}}-\Sigma_{x,\text{ROM}}||_F}{||\Sigma_{x,\text{FOM}}||_F},
    \label{eq:errorcov}
\end{equation}
where $||.||_F$ denotes the Frobenius norm. Eq.~\ref{eq:errorcov} introduces a relative error measure that shows how far the computed covariance matrices differ from one another. In order to compute the covariance matrix of $\mathbf{x}$, Eq.~\ref{eq:covsvd} is solved with the created ROM. With the obtained ROM low-rank factors the covariance of $\mathbf{x}$ is finally computed and the error measure can be evaluated. Tab.~\ref{tab:coverr} shows the error for certain selected frequencies. It should be mentioned here, that not all errors can be shown, since the computation of $\Sigma_{x,\text{FOM}}$ is too computationally expensive for every single frequency step. Therefore, only some exemplary frequencies are chosen throughout the examined frequency domain that highlight the overall trend of the computed error.
\begin{table}[h!]
    \centering
    \caption{Error of the approximated covariance matrices over frequency.}
    \begin{tabular}{c|c}
         Frequency& Error \\
         \hline
         $16$~Hz&$2.24\times 10^{-4}$ \\
         $32$~Hz&$2.32\times 10^{-6}$ \\
         $114$~Hz& $3.45\times 10^{-6}$ \\
         $414$~Hz& $1.82\times 10^{-4}$ \\
         $460$~Hz& $9.20\times 10^{-3}$ \\
    \end{tabular}
    \label{tab:coverr}
\end{table}
Tab.~\ref{tab:coverr} shows a relative error, hence, the approximations of the covariance matrix through low-rank factors obtained by using a ROM works quite well. The relative error is less than $1\%$ across the examined frequency domain. Two of the examined frequencies are the respective expansion points of the ROM, which means the error should be comparably low, since the ROM is exact at the expansion points. The interesting part, however, is that even away from the expansion points the covariance matrix gets estimated quite well by the proposed method, since in a low frequency regime the error is even lower. Still, it can also be observed that the error is increasing with frequency. This is due to the fact that the ROM also gets worse with increasing frequency. The higher modal density makes it tougher to approximate the system's behavior and therefore more expansion points could be introduced for a gain in accuracy. This always comes with an increase in ROM size, therefore, one has to consider computational effort versus accuracy when introducing more expansion points into the ROM algorithm. Nonetheless, an error of $9.20\times 10^{-3}$ for $460$~Hz is still seen as sufficiently small, showcasing that the proposed method works in estimating covariance matrices through ROM low-rank factors.\\
The methodology introduced in this contribution shows, that a combination of low-rank approximation and MOR works in a vibroaocustic setting with stochastic loads. However, so far the computational effort has been omitted from the discussion, which is the topic of the following paragraph. All computations were conducted on the same system under the same circumstances, namely with a AMD Ryzen Threadripper 3960X 24-Core Processor, Base clock: 3.8 GHz, number of threads: 48 (24x2) , OS: Ubuntu 20.04.6 LTS, and 126 GB RAM. Therefore, the solving times have to be seen with regards to the machine used. All the codes were implemented in MATLAB. Tab.~\ref{tab:timings} shows a comparison of the computational effort of the proposed method and the FOM solution times.

\begin{table}[h!]
    \centering
    \caption{Solution times per frequency step for different computation methods.}
    \begin{tabular}{c|c|c|c}
         Operation&ROM  &FOM  &FOM low-rank \\
         \hline
         Compute $\mathbf{V}$&$324$~s (offline)  & - & -\\
         SVD& $2$~s & - &$2$~s \\
         Low-rank factors& $40$~s  &-  &$264$~s \\
         Assembly& $126$~s & $9115$~s & $126$~s\\
         \hline
         Total&$168$~s ($492$~s)  &$9115$~s  & $392~s$\\
    \end{tabular}
    \label{tab:timings}
\end{table}
Tab.~\ref{tab:timings} shows the comparisons of solution times for the proposed method (ROM) and the straight forward way of computing the covariance matrices (FOM), as well as computing the covariance matrices of the FOM with a low-rank approximation (FOM low-rank). The solving for the mean is included in the computation of the low-rank factors, since the system has to be solved for only one more RHS here. It is important to notice that the times are per frequency step, meaning that for each frequency this procedure has to be repeated. Another important aspect is the differentiation between off- and online phase in the ROM column. The projection matrix $\mathbf{V}$ only has to be computed once, since it is not frequency dependent, meaning the first frequency step will have a time of $492$~s, while all the following frequency steps will only take $168$~s. Hence, the MOR approach would not be beneficial for small frequency bands, however, since we are evaluating the model at many frequency points, the proposed method will be more efficient with regards to computational time than the original method. This can be seen by comparing the total ROM time to the FOM time, which is significantly larger. Finally, the low-rank approach could also be applied to the FOM directly. However, Tab.~\ref{tab:timings} shows that after two frequency steps, the proposed method is already faster than the FOM low-rank approach.\\

\section{Conclusion}
This contribution successfully showcases the application of a combined low-rank MOR approach to a vibroacoustic benchmark model. The plate cavity system is excited by a TBL which is subject to uncertainties due to a random phase being added. For this setting, standard uncertainty quantification can be computationally expensive. We have targeted the computation of the solution's mean and covariance matrix at every frequency step. To this end, we applied a low-rank approximation to the covariance matrix of the excitation to propagate the so-called low-rank factors through the system of equations, so that the covariance matrix of the vector of unknowns can be built. Still, many DOFs make the evaluation challenging with regards to computational effort, since the computational time, especially for finer discretized models might still be high. Therefore, we also applied a MOR method to reduce the original system's size by a significant amount. It was possible to accurately predict the transfer function of the system, as well as the covariance matrices by only creating one projection matrix, which had the capability to estimate both mean and low-rank factors. Hence, the computational cost, quantified by solving times, could be reduced by a significant amount, therefore enabling many evaluations of the system, making frequency sweeps and uncertainty quantification possible.\\
We could show that the method works well for models that are moderate in size and that it represents a promising approach towards uncertainty quantification in vibroacoustics. It still remains to be seen if a more realistic cabin noise model will also show the same kind of results. Furthermore, material uncertainties could also be introduced into the model, which will be an interesting task for future research.\\

\section{Acknowledgements}
We acknowledge the funding by the Deutsche Forschungsgemeinschaft (DFG, German Research Foundation) under Germany´s Excellence Strategy – EXC 2163/1 – Sustainable and Energy Efficient
Aviation – Project-ID 390881007.
\bibliographystyle{plain}
\bibliography{main}
\end{document}